\documentclass[showpacs,amsmath,amssymb,prb,superscriptaddress,twocolumn]{revtex4}
\usepackage[T1]{fontenc}
\usepackage{amssymb,amsmath}
\usepackage{times}
\usepackage{graphicx}
\usepackage{wasysym}
\usepackage{subfigure}
\usepackage{bm}% bold math

\begin{document}

\title{Self-consistent calculations of loss compensated fishnet metamaterials}

\author{A.~Fang}
\affiliation{Department of Physics and Astronomy and Ames Laboratory,
             Iowa State University, Ames, Iowa 50011, U.S.A.}

\author{Th.~Koschny}
\affiliation{Department of Physics and Astronomy and Ames Laboratory,
             Iowa State University, Ames, Iowa 50011, U.S.A.}
\affiliation{Department of Materials Science and Technology and Institute of Electronic Structure and Laser, FORTH,
              University of Crete, 71110 Heraklion, Crete, Greece}

%\author{M.~Wegener}
%\affiliation{Institut f\"ur Angewandte Physik and DFG-Center for Functional Nanostructures (CFN),
%             Universit\"at Karlsruhe (TH), D-76128 Karlsruhe, Germany}

\author{C.~M.~Soukoulis}
\affiliation{Department of Physics and Astronomy and Ames Laboratory,
             Iowa State University, Ames, Iowa 50011, U.S.A.}
\affiliation{Department of Materials Science and Technology and Institute of Electronic Structure and Laser, FORTH,
              University of Crete, 71110 Heraklion, Crete, Greece}

\date{\today}

%%% ABSTRACT

\begin{abstract}
We present a computational approach, allowing for a self-consistent treatment of three-dimensional (3D) fishnet metamaterial coupled to a gain material incorporated into the nanostructure. We show numerically that one can compensate the losses by incorporating gain material inside the fishnet structure. The pump rate needed to compensate the loss is much smaller than the bulk gain and the figure of merit ($\mathrm {FOM} = |\mathrm {Re}(n)/\mathrm {Im}(n)|$) increases dramatically with the pump rate. Transmission, reflection, and absorption data, as well as the retrieved effective parameters, are presented for the fishnet structure with and without gain material. Kramers-Kronig relations of the effective parameters are in excellent agreement with the retrieved results with gain.
\end{abstract}

% 41.20.Jb  Electromagnetic wave propagation; radiowave propagation
% 42.25.Bs  Wave propagation, transmission and absorption
% 42.70.Qs  Photonic bandgap materials
% 73.20.Mf  Collective excitations
% 78.20.Ci  Optical constants (including refractive index, complex dielectric constant, T, R, A, emissivity)


\maketitle
The field of metamaterials has seen spectacular experimental progress in recent years. \cite {1a,1b,1,2,3,3a} Yet, losses are orders of magnitude too large for the envisioned applications, such as, e.g., perfect lenses, \cite {3b} and invisibility cloaking.\cite {3c} For some applications, such as perfect absorbers, \cite {3d} the loss is not a problem. There are two basic approaches to address the loss. One approach to reduce the metamaterial losses to some extent is by geometric tailoring of the metamaterial designs. \cite {4, 5,6,7} An efficient method to geometrically reduce the losses in metamaterials is by increasing the inductance, $L$, to the capacitance, $C$, ratio, \cite {4} and avoid corners and sharp edges in metamaterials. \cite {7} Another method to reduce losses is to move the real part of the negative index of refraction, $n$, away from the maximum of $\mathrm {Im}(n)$ (close to the resonance) \cite {5,6} by strongly coupled metamaterials. The second approach is to incorporate gain material into metamaterial designs. One important issue is not to assume the metamaterial layer and the gain medium layer are independent from one another. \cite {7a,7b,8,9,10,11,12,13} So, there is a need for self-consistent calculations \cite {14,15,16} for incorporating gain materials into realistic metamaterials. The need for self-consistent calculations stems from the fact that increasing the gain in the metamaterial, the metamaterial properties change, which, in turn, changes the coupling to the gain medium until a steady state is reached. 

A time-domain self-consistent calculation of gain in two-dimensional (2D) magnetic metamaterial has been recently reported. \cite {15,16} It has been demonstrated \cite {15} that the losses of the magnetic susceptibility, $\mu$, of the split-ring resonator (SRR) can be compensated by the gain material. So, these self-consistent calculations \cite {14,15,16} will guide the experimentalists to seek new 3D metamaterial designs where the gain medium will give an effective gain much higher than its bulk counterpart and reduce the losses. This large value of gain is due to the strong local-field enhancement inside  the metamaterial design. With regard to experiments to reduce losses in metamaterials, one needs to use semiconductor gain (quantum dots or wells) and not use dye molecules, \cite {17} which photo-bleach rapidly. Semiconductor gain enables long-term use and can be conceptually pumped by electrical injection. This is crucial, as applications based on optically-pumped structures do not appear to be realistic in the long run. However, to check if losses in metamaterials can be reduced experimentally, one can try exploratory experiments under conditions of optical pumping.

In this paper, we apply a detailed 3D computational model to study the optical response of the fishnet structure with gain medium embedded in the structure. We find that complete loss-compensation is possible with gain medium and the $\mathrm {FOM}$ increases dramatically with the pump rate. \cite {18} A similar time-domain calculation of gain in fishnet metamaterials has been recently reported. \cite {19} The gain is described by a generic four-level atomic system, which tracks fields and occupation numbers at each point in space, taking into account energy exchange between atoms and fields, electronic pumping, and non-radiative decays. \cite{20} An external mechanism pumps electrons from the ground state level, $N_0$, to the third level, $N_3$, at a certain pump rate, $\Gamma_\mathrm{pump}$, proportional to the optical pumping intensity in an experiment. After a short lifetime, $\tau_{32}$, electrons transfer non-radiatively into the metastable second level, $N_2$.  The second level ($N_2$) and the first level ($N_1$) are called the upper and lower lasing levels. Electrons can be transferred from the upper to the lower lasing level by spontaneous and stimulated emission. At last, electrons transfer quickly and non-radiatively from the first level ($N_1$) to the ground state level ($N_0$). The lifetimes and energies of the upper and lower lasing levels are $\tau_{21},\ E_2$ and $\tau_{10},\ E_1$, respectively.  The center frequency of the radiation is $\omega_a=(E_2-E_1)/\hbar$, chosen to equal $2\pi \times 1.5\times 10^{14}\,\mathrm{Hz}$.  The parameters, $\tau_{32}$, $\tau_{21}$, and $\tau_{10}$, are chosen $5\times10^{-14}$, $5\times10^{-12}$, and $5\times10^{-14}\,\mathrm{s}$, respectively.  The total electron density, $N_0(t=0) = N_0(t) + N_1(t) + N_2(t) + N_3(t) = 5.0\times 10^{23}\,\mathrm {/m^3}$, and the pump rate, $\Gamma_\mathrm{pump}$, is an external parameter. These gain parameters are chosen to overlap with the resonance of the fishnet. The time-dependent Maxwell equations are given by $\nabla\times \mathbf {E} = -\partial \mathbf {B}/\partial t$ and $\nabla\times \mathbf {H} = \varepsilon \varepsilon_o \partial \mathbf {E}/\partial t + \partial \mathbf {P}/\partial t$, where $\mathbf {B}=\mu\mu_o \mathbf {H}$ and $\mathbf {P}$ is the dispersive electric polarization density from which the amplification and gain can be obtained. Following the single electron case, we can show \cite{20} the polarization density $\mathbf {P}(\mathbf {r},t)$ in the presence of an electric field obeys locally the following equation of motion,
\begin{equation}
 \label{Eqn:1}
 \frac{\partial^2 \mathbf {P}(t)}{\partial t^2} +
 \Gamma_a\frac{\partial \mathbf {P}(t)}{\partial t} +
 \omega_a^2 \mathbf {P}(t) \ =\
 -\sigma_a \Delta N(t) \mathbf {E}(t),
\end{equation}
where $\Gamma_a$ is the linewidth of the atomic transition $\omega_a$ and is equal to $2\pi \times 20 \times 10^{12} \,\mathrm{Hz}$. The factor, $\Delta N(\mathbf {r},t) = N_2(\mathbf {r},t) - N_1(\mathbf {r},t)$, is the population inversion that drives the polarization, and $\sigma_a$ is the coupling strength of $\mathbf {P}$ to the external electric field and its value is taken to be $10^{-4}\,\mathrm {C^2/kg}$. It follows \cite {20} from Eq.~\ref{Eqn:1} that the amplification line shape is Lorentzian and homogeneously broadened. The occupation numbers at each spatial point vary according to
\begin{subequations} 
 \begin{align}
 \frac{\partial N_3}{\partial t} &= \Gamma_\mathrm{pump}\, N_0 -
                                    \frac{N_3}{\tau_{32}} \\
 \frac{\partial N_2}{\partial t} &= \frac{N_3}{\tau_{32}} +
                                    \frac{1}{\hbar\omega_a} \mathbf {E}\cdot\frac{\partial \mathbf {P}}{\partial t}  -
                                    \frac{N_2}{\tau_{21}} \\
 \frac{\partial N_1}{\partial t} &= \frac{N_2}{\tau_{21}} -
                                    \frac{1}{\hbar\omega_a} \mathbf {E}\cdot\frac{\partial \mathbf {P}}{\partial t}  -
                                    \frac{N_1}{\tau_{10}}  \\
 \frac{\partial N_0}{\partial t} &= \frac{N_1}{\tau_{10}} -
                                    \Gamma_\mathrm{pump}\, N_0, 
 \end{align}
\end{subequations}
where $\frac{1}{\hbar\omega_a}\mathbf {E}\cdot\frac{\partial \mathbf {P}}{\partial t}$ is the induced radiation rate or excitation rate depending on its sign.

To solve the behavior of the active materials in the electromagnetic fields numerically, the finite-difference time-domain (FDTD) technique is utilized. \cite{21} In the FDTD calculations, the discrete time and space steps are chosen to be $\Delta t=8.0 \times 10^{-18}\,\mathrm{s}$ and $\Delta x=5.0\times10^{-9}\,\mathrm{m}$. The initial condition is that
all the electrons are in the ground state, so there is no field, no polarization, and no spontaneous emission. Then, the electrons are pumped from $N_0$ to $N_3$ (then relaxing to $N_2$) with a constant pump rate, $\Gamma_\mathrm{pump}$. The system begins to evolve according to the system of equations above.

\begin{figure}
 \centering
  \includegraphics[width=0.5\textwidth]{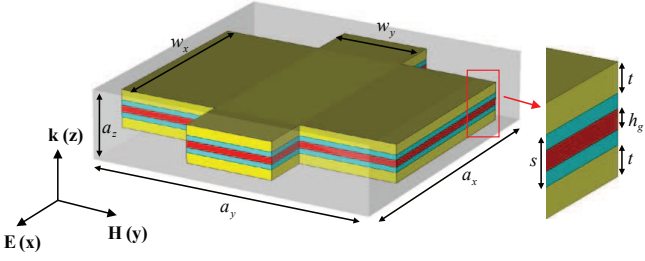}
 \caption {%
  (Color online)
    Schematic of the unit cell of the fishnet structure with the parameters marked on it. 
    The geometric parameters are $a_x = a_y = 860\,\mathrm {nm}$, $a_z = 200 \,\mathrm {nm}$, $w_x = 565\,\mathrm {nm}$, 
    $w_y = 265\,\mathrm {nm}$, $s = 50\,\mathrm {nm}$, $t = 30\,\mathrm {nm}$ and $h_g = 20\,\mathrm {nm}$. 
    The thicknesses of the metal (silver) and gain layer are $t$ and $h_g$, respectively, and the dielectric constant of the spacer, $\mathrm {MgF_2}$, is 1.9. 
    These parameters were used on simulations \cite {6} and experiments. \cite {5} 
  }
  \label{fig1}
\end{figure}

In Fig.~1, we show the unit cell of the fishnet structure. The size of the unit cell along the propagation direction is $a_z$. $a_z$ is larger than the sum of the thickness of the metallic and the dielectric layers $2t+s$, where $t$ and $s$ are thicknesses of the metal and the dielectric layers, respectively. Notice the propagation direction is perpendicular to the plane of the fishnet with the electric and magnetic fields along the $x$ and $y$ directions, respectively. All retrieved effective parameters are for this particular incident direction and field polarization. Consider two configurations, one without gain and one with gain (see inset of Fig.~1). In the configuration with gain, we have introduced two thin dielectric layers of thickness, $(s-h_g)/2$, close to the metallic structure, so the gain medium will not be close to the metal, then quenching will be avoided. The dimensions of the fishnet structure \cite {5,6} are chosen such that the magnetic resonance wavelength at $\lambda = 2000\,\mathrm {nm}$, which can overlap with the peak of the emission of the gain material. The full width at half maximum (FWHM) of the gain material is $20\,\mathrm {THz}$ and the pump rate, $\Gamma_{\mathrm {pump}}$, changes from 0 to $6.9\times 10^8\,\mathrm {s^{-1}}$.

\begin{figure}
 \centering
  \subfigure{ 
   \includegraphics[width=0.4\textwidth]{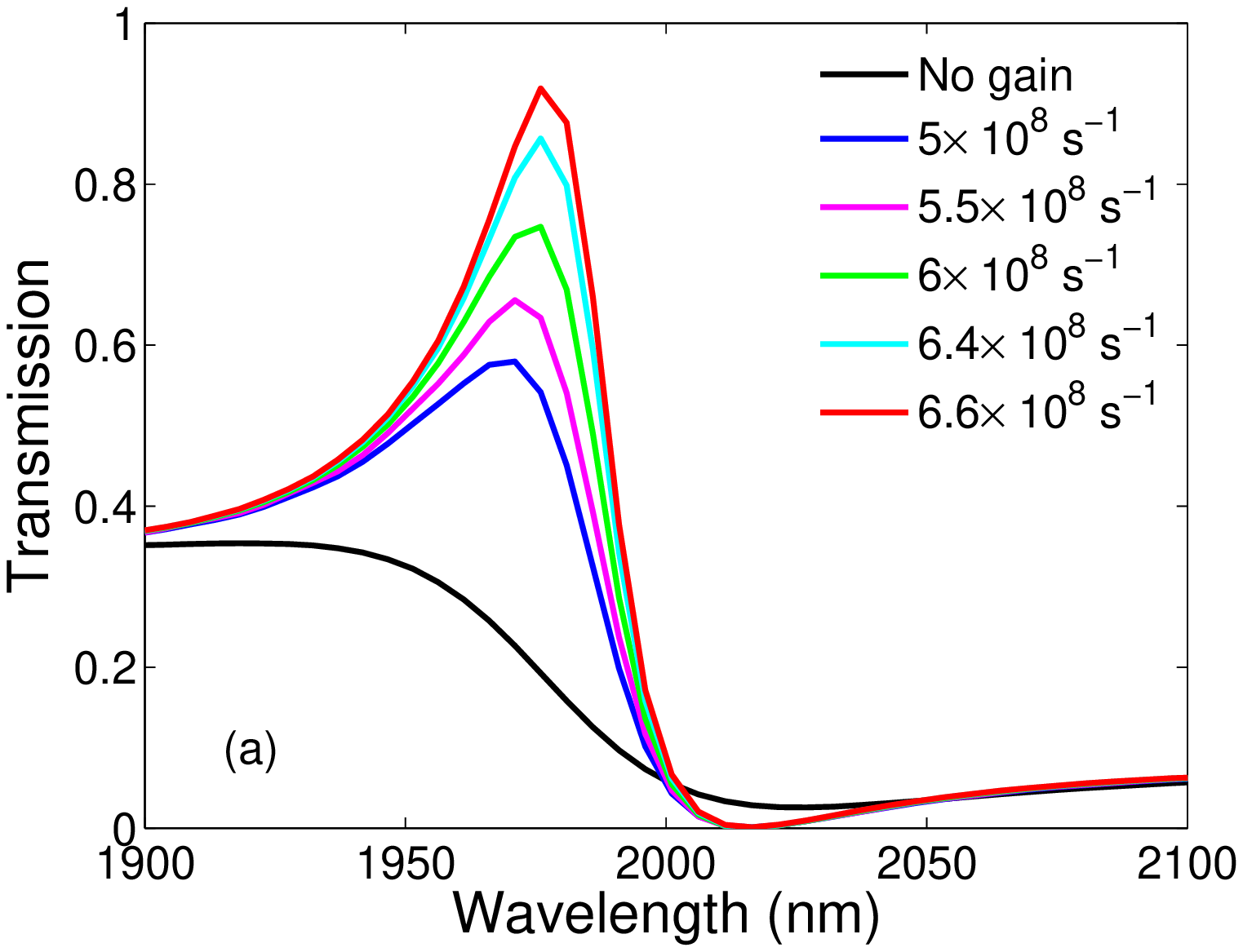}
   }
 \centering
  \subfigure{
   \includegraphics[width=0.4\textwidth]{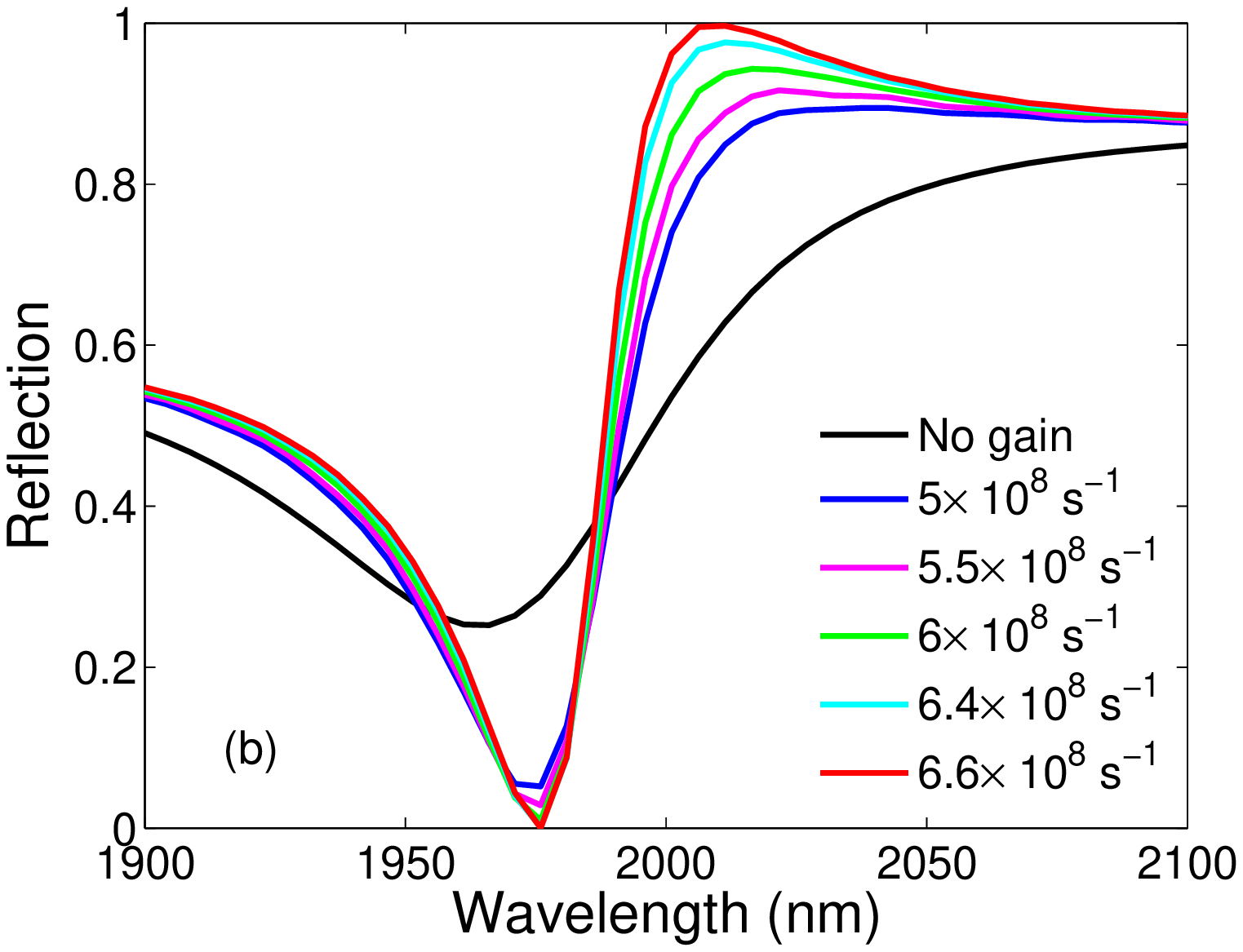}
  }
 \centering
  \subfigure{
   \includegraphics[width=0.4\textwidth]{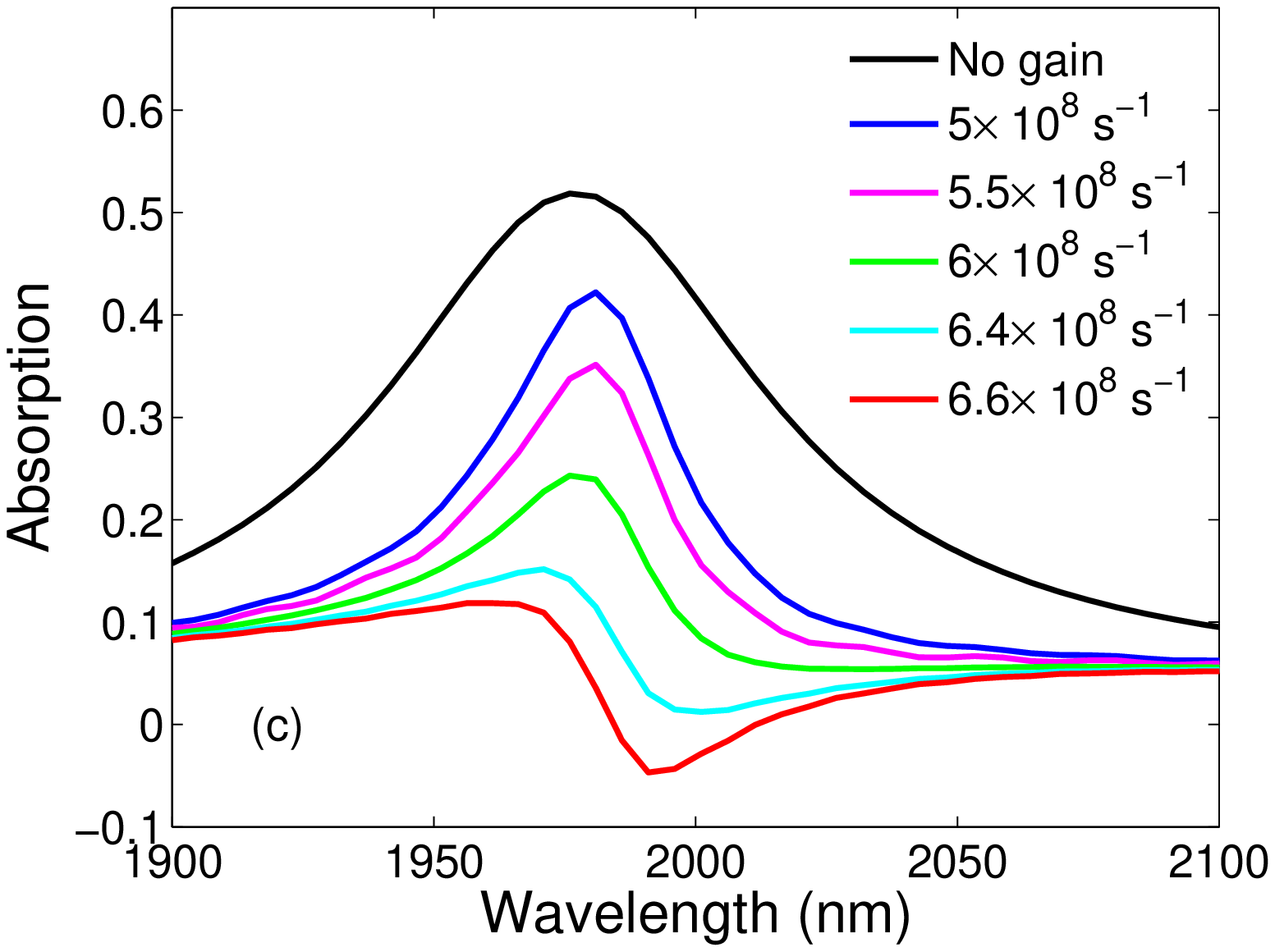}
  }
 \caption{%
  (Color online)
  The transmission (a), reflection (b) and absorption (c) as a function of wavelength for different pump rates. }
 \label{fig2}
\end{figure}
\begin{figure}
 \centering
  \subfigure{ 
   \includegraphics[width=0.4\textwidth]{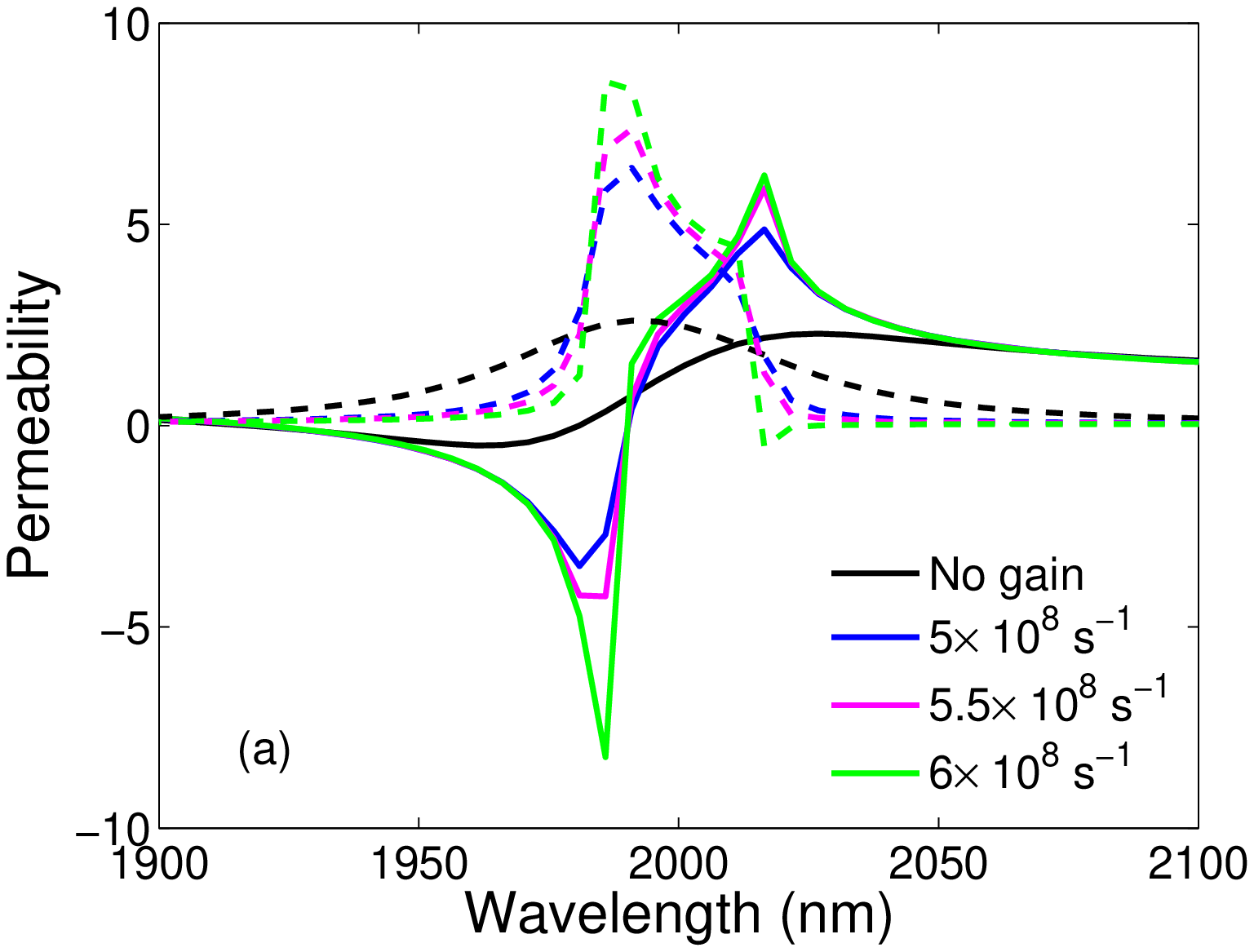}
   }
 \centering
  \subfigure{
   \includegraphics[width=0.4\textwidth]{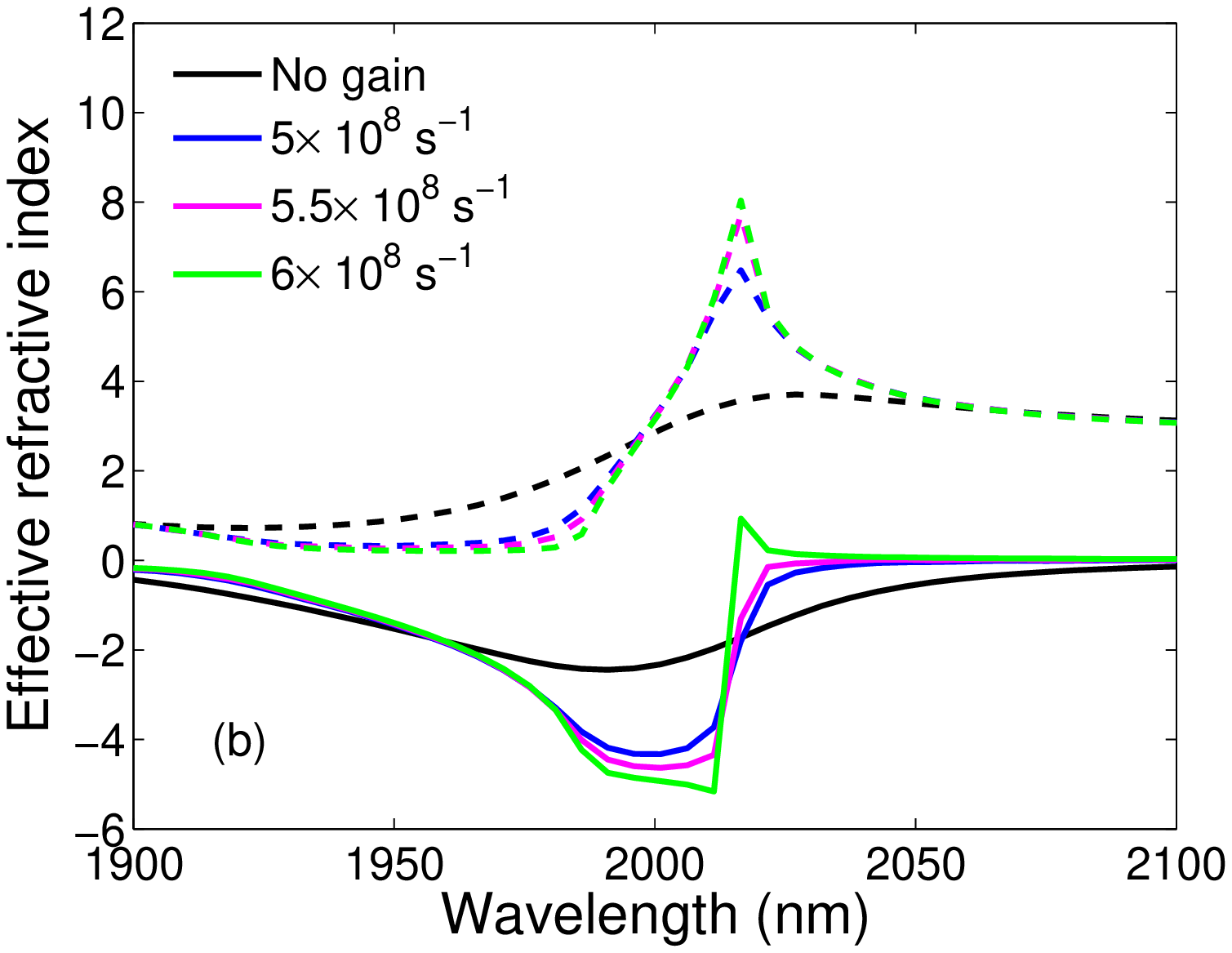}
  }
 \centering
  \subfigure{
   \includegraphics[width=0.4\textwidth]{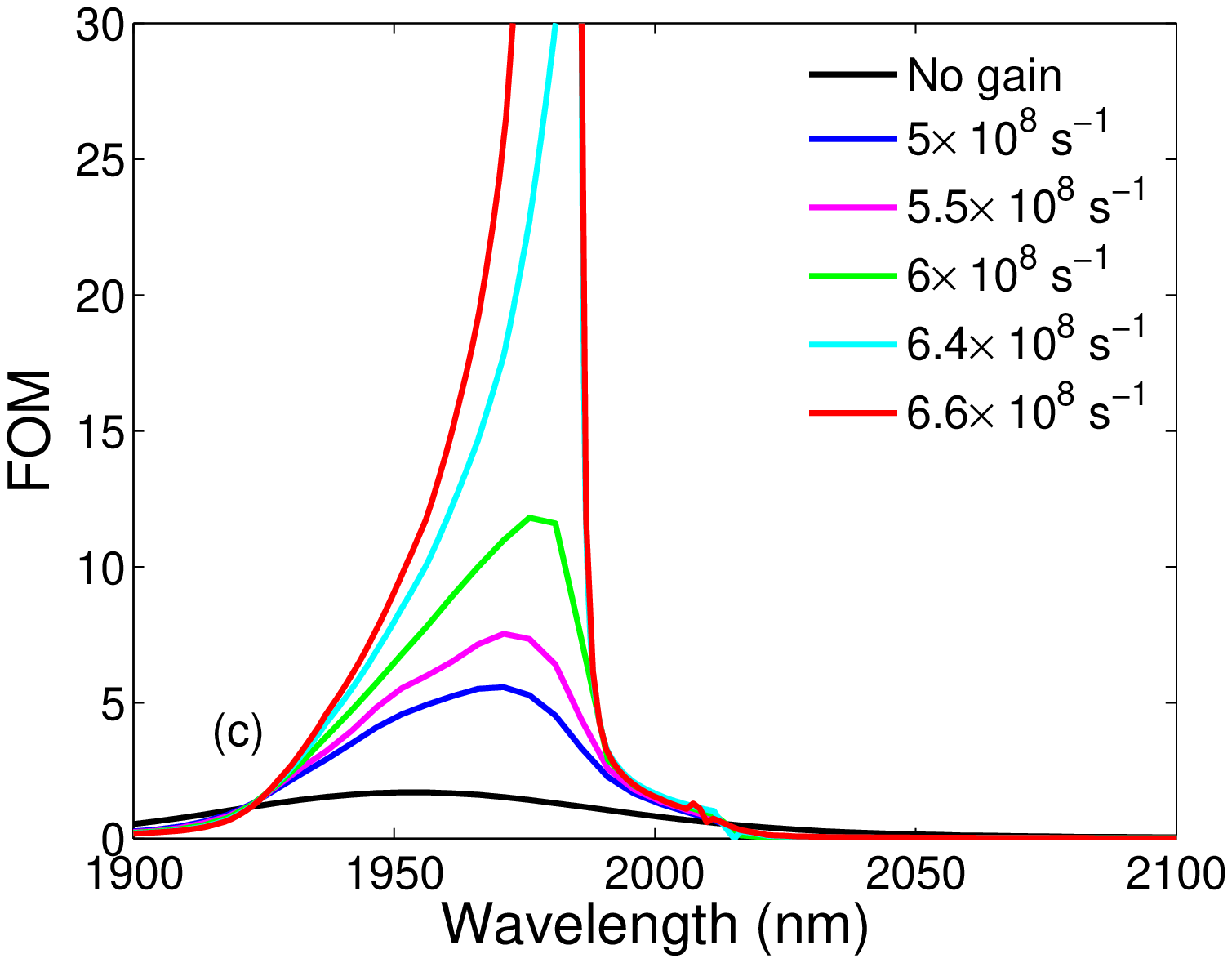}
  }
 \caption{%
  (Color online)
    The retrieved results for the real (solid lines) and the imaginary (dashed lines) parts of
  (a) the effective permeability, $\mu$, and 
  (b) the corresponding effective index of refraction, $n$, without and with gain for different pump rates. (c) The figure-of-merit ($\mathrm {FOM}$) as a function of wavelength
  for different pump rates. 
  The gain bandwidth is $20\,\mathrm {THz}$.
  }
 \label{fig3}
\end{figure}
In Fig.~2, we plot the transmission, $T=|t|^2$, (2a), reflection, $R=|r|^2$, (2b) and absorption, $A=1-T-R$, (2c) versus wavelength for different pump rates ($t$ and $r$ are the transmission and reflection amplitudes, respectively). Notice the wavelength dependence of $T$ and $R$ for different pump rates away from the resonance wavelength, $\lambda = 2000\,\mathrm {nm}$, are the same. Below the resonance wavelength, $T$ increases with the pump rate and above the resonance wavelength, $T$ decreases with the pump rate. The reflection $R$, below the resonance wavelength, decreases with the pump rate and above the resonance wavelength, it increases with the pump rate. Notice in Fig.~2a, $T$ without gain has a very weak resonance, and once the pump rate increases, the transmission clearly shows the resonance behavior. The same can be seen in the experiments that can use gain materials to compensate the losses in fishnet metamaterials. In Fig.~2c, we plot the absorption, $A$, as a function of wavelength for different pump rates. Notice, as we increase the pump rate, the absorption decreases and finally at the pump rate of $6.6\times 10^9\,\mathrm {s^{-1}}$, the gain overcompensates the losses and the absorption becomes negative. In Fig.~3a, we plot the retrieved results of the real and imaginary parts of the magnetic permeability, $\mu$, with and without gain, for normal incidence and the particular field polarization in Fig.~1. As gain increases, the $\mathrm {Re}(\mu)$ becomes steeper at the resonance wavelength and the $\mathrm {Im}(\mu)$ becomes much narrower when increasing the pump rate and the losses are compensated by the gain material. In Fig.~3b, we plot the retrieved results for the effective index of refraction $n$, with and without gain. The $\mathrm {Re}(n)$ becomes more negative after gain is introduced and the $\mathrm {Im}(n)$ also drops significantly close to the resonance. At $\lambda = 1976\,\mathrm {nm}$, the $\mathrm {Re}(n)$ changes from -2.25 to -2.82 with a pump rate of $5.0\times 10^8\,\mathrm {s^{-1}}$ and the $\mathrm {Im}(n)$ drops from 1.58 to 0.54 (Fig.~3b). In Fig.~3c, we plot the $\mathrm {FOM}$ versus the wavelength for different pump rates. Notice, the  $\mathrm {FOM}$ becomes very large (of order of $10^2$) with the pump rates. Comparing $\mathrm {Im}(n)$ slightly below the resonance at $\lambda = 1976\,\mathrm {nm}$, we find the effective extinction coefficient $\alpha = \frac {\omega}{c}\mathrm {Im}(n) \approx 5.0\times 10^4\,\mathrm {cm^{-1}}$ without  gain and $\alpha \approx 1.7\times 10^{4}\,\mathrm {cm^{-1}}$ with gain ($\Gamma_{\mathrm {pump}} = 5.0\times 10^8\,\mathrm {s^{-1}}$). Hence, an effective amplification of $\alpha = -3.3\times10^4\,\mathrm {cm^{-1}}$. This is much larger (of the order of $30$) than the expected amplification $\alpha \approx -1.3\times 10^3\,\mathrm {cm^{-1}}$ for the gain material at the given pump rate of $5.0\times 10^8\,\mathrm {s^{-1}}$. The difference can be explained by the field enhancement in the fishnet metamaterial.

\begin{figure*}
 \centering
  \subfigure{ 
   \includegraphics[width=0.4\textwidth]{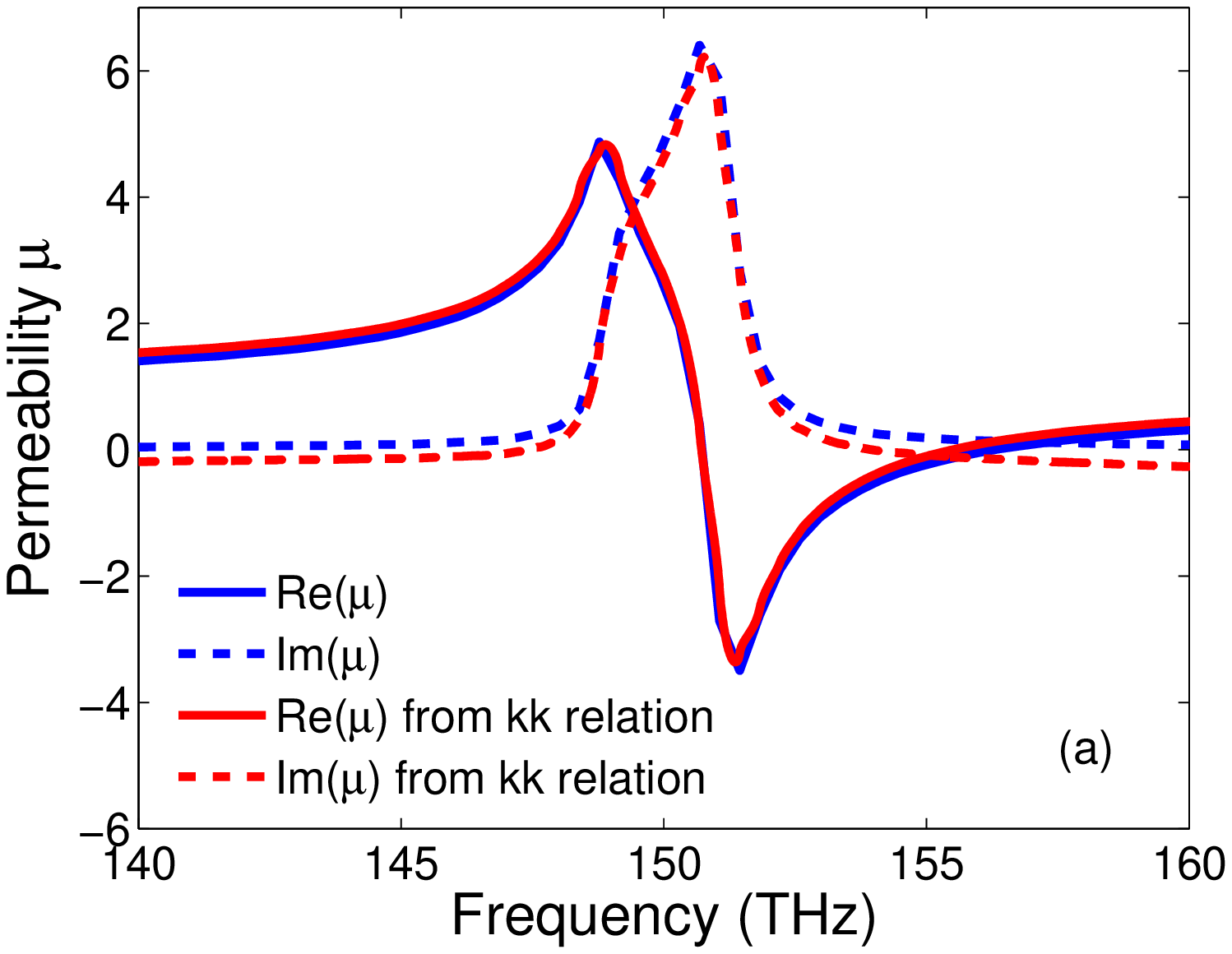}
   }
 \centering
  \subfigure{
   \includegraphics[width=0.4\textwidth]{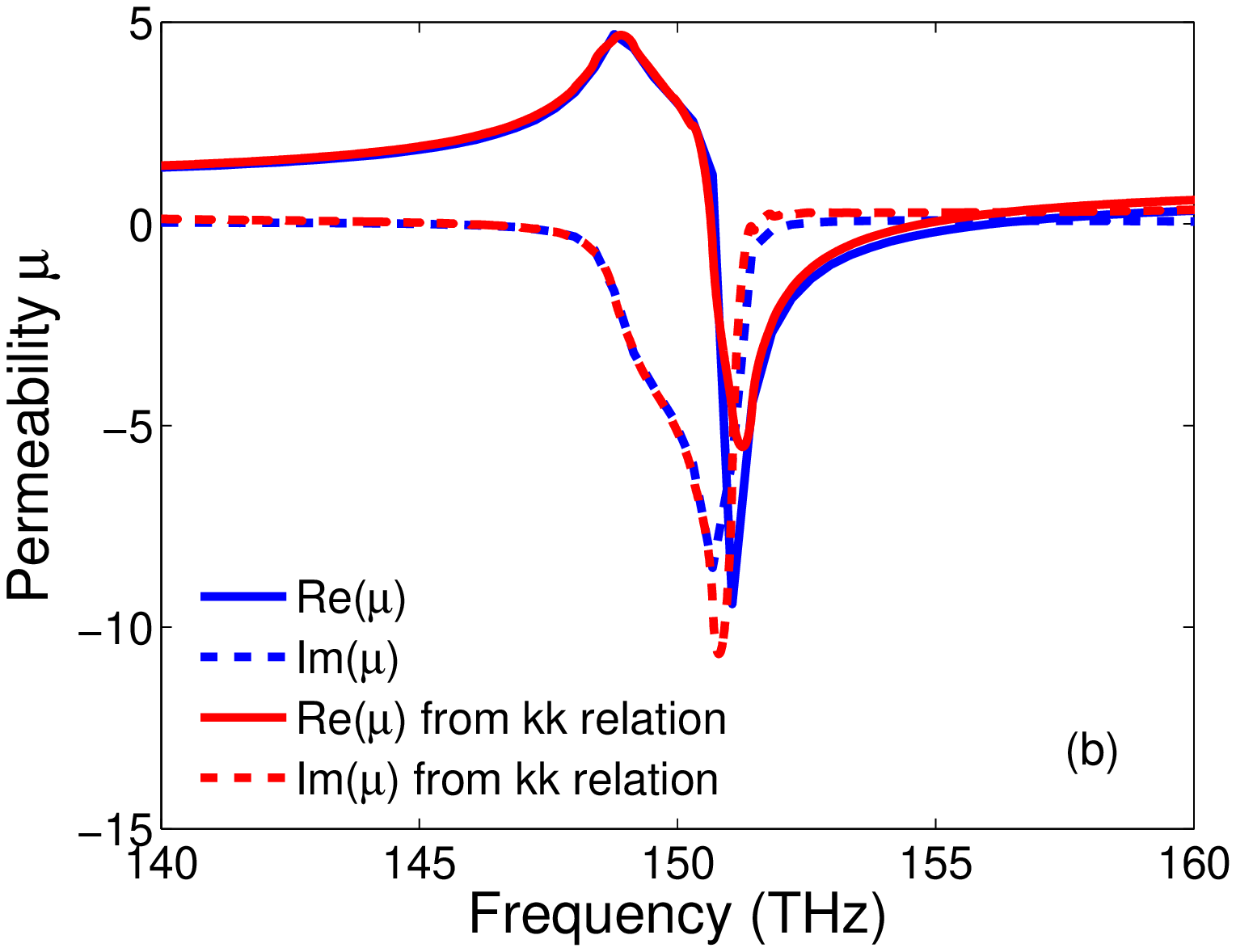}
  }
  \caption{%
  (Color online)
    The real and imaginary parts of the retrieved effective permeability, $\mu$, and the results from the Kramers-Kronig relations for pump rates (a) $\Gamma_{\mathrm {pump}}=5.0\times10^8\,\mathrm {s^{-1}}$ (below compensation) and (b) $\Gamma_{\mathrm {pump}}=6.9\times10^8\,\mathrm {s^{-1}}$ (overcompensated).
 }
 \label{fig4}
\end{figure*}
There are theoretical debates \cite {22,23,24} if it is possible to obtain low loss metamaterials with negative refractive index, $n$. They have used the Kramers-Kronig (KK) relations and we would like to verify that Kramers-Kronig relations work with and without gain. In addition, we need to compare the numerically-retrieved effective permeability, $\mu$, shown in Fig.~3a, with the calculation of $\mu$, based on the Kramers-Kronig relations. In Fig.~4a, we plot the real and imaginary parts of the effective permeability, $\mu$, and the results from the Kramers-Kronig relations for the pump rate of $\Gamma_{\mathrm {pump}}=5.0\times 10^8\,\mathrm {s^{-1}}$. The excellent agreement between the results obtained from the standard retrieval method \cite {25} (a special case of anisotropic retrieval method \cite {26} for normal incidence) and the Kramers-Kronig approach verifies KK relations work for metamaterials coupled with gain too. KK relations are valid for passive materials and the deduction of KK relations is based on the assumption that the response function of materials is analytical in the upper complex plane of frequency. For strong active materials, there are poles in the upper plane and one must modify the KK relations by reversing the signs of KK relations. \cite {24} In Fig.~4b, we plot the $\mathrm {Re}(\mu)$ and $\mathrm {Im}(\mu)$ obtained by the retrieval method and by the modified KK relations for the pump rate $\Gamma_{\mathrm {pump}} = 6.9\times 10^8\,\mathrm {s^{-1}}$. For this pump rate the $\mathrm {Im}(\mu)$ becomes negative and we have overcompensated at the resonance frequency of $150\,\mathrm {THz}$. As one can see from Fig.~4b, KK relations work well for strongly active materials.

In conclusion, we have proposed and numerically solved a self-consistent model incorporating gain in the 3D fishnet dispersive metamaterial. We show numerically that one can compensate the losses of the fishnet metamaterial. We have presented results for $T$, $R$, and $A$ without and with gain for different pump rates. Once the pump rate increases, both $T$ and $R$ show a resonance behavior. We have retrieved the effective parameters for different pump rates and the losses are compensated with gain. Kramers-Kronig relations of the effective parameter are in excellent agreement with the retrieved results with gain. The figure-of-merit ($\mathrm {FOM}$) with gain increases dramatically and the pump rate needed to compensate the loss is much smaller than the bulk gain. This aspect is due to the strong local-field enhancement inside the fishnet structure.

We thank M.~Wegener for useful discussions. Work at Ames Laboratory was supported by the Department of Energy (Basic Energy Sciences) under Contract No.~DE-AC02-07CH11358. This work was partially supported by the European Community FET project PHOME (Contract No.~213390) and by Laboratory-Directed Research and Development Program at Sandia National Laboratories.

\bibliographystyle{apsrev}

\end{document}